# Theory of a Continuous $H_{c2}$ Normal-to-Superconducting Transition

Leo Radzihovsky

*The James Franck Institute and Physics Department, University of Chicago, 5640 South Ellis Avenue, Chicago, IL 60637*

*Institute for Theoretical Physics, University of California, Santa Barbara, CA 93106-4030*

(March 28, 1995)

I study the $H_{c2}$ transition within the Ginzburg-Landau model, with $m$-component order parameter $\psi_i$. I find a renormalized fixed point free energy, exact in $m \to \infty$ limit, suggestive of a 2nd-order transition in contrast to a general belief of a 1st-order transition. The thermal fluctuations for $H \neq 0$ force one to consider an infinite set of marginally relevant operators for $d < d_{uc} = 6$. I find $d_{lc} = 4$, predicting that the ODLRO does not survive thermal fluctuations in $d = 2, 3$. The result is a solution to a critical fixed point that was found to be inaccessible within $\epsilon = 6 - d$-expansion, previously considered in E.Brezin, D.R.Nelson, A.Thiaville, Phys.Rev.B **31**, 7124 (1985), and was interpreted as a 1st-order transition.

64.60.Fr, 74.20.D

The nature of the Normal-to-Superconducting (NS) transition is substantially more challenging than the neutral superfluid transition and has remained controversial for many years. While there is now finally a general consensus on the nature of the *zero* field NS transition [1–3], the *finite* field NS transition is much less understood. As discovered by Abrikosov [4], in mean-field theory, upon lowering $T$ the NS transition takes place at $H_{c2}$ ($T_c(H)$) and the superconducting order parameter develops a triangular array of flux-line vortices, a transition believed to be describable by the Ginzburg-Landau free energy

$$F[\psi, \vec{A}] = \int d^d r \left[ |(\vec{\nabla} + iq_o\vec{A})\psi|^2 \right.$$
$$\left. + r_o|\psi|^2 + \frac{1}{2}g_o(|\psi|^2)^2 + \frac{1}{8\pi\mu_o}(\vec{\nabla} \times \vec{A} - \vec{H})^2 \right], \quad (1)$$

where $r_o \sim (T - T_c)/T_c$, $q_o = 2e/\hbar c$, and $\mu_o$ is magnetic permeability of the normal metal and is close to unity. Although it is generally believed that in mean-field the transition is of second order, even this theory is not on a completely firm ground. The issue is that at the transition two apparently unrelated symmetries are supposed to be simultaneously broken: the translational symmetry is broken by the formation of the Abrikosov lattice accompanied by breaking of $U(1)$ gauge symmetry leading to phase coherence and superconductivity. While it is easy to see the diverging uniform susceptibility ($k = 0$), associated with the formation of phase coherence, no corresponding divergent susceptibility for the formation of a lattice (much less a specific triangular lattice) had been demonstrated. Following Abrikosov, what is generally demonstrated is that a periodic solution (e.g. triangular) is lower in energy, globally, than a lattice of another symmetry (e.g. the square one).

The gaussian fluctuation corrections only to the specific heat around $H_{c2}$ (valid only outside critical region) were first studied in the pioneering work by Lee and Shenoy [5], followed by a more sophisticated self-consistent method and expansions in the range of the nonlinear interaction [6]. These results for the specific heat are consistent with my more general and controlled analysis of the full critical behavior.

Although a form of freezing into a 3d lattice takes place, as was first emphasized by Brezin, Nelson and Thiaville (BNT) [7], Landau's general argument for the 1st-order freezing transition is circumvented by the fact that in mean-field both of the symmetries are broken simultaneously and the amplitude of $\psi$ is small, preventing the density from jumping discontinuously. Since it is likely that this will no longer be the case, once thermal fluctuations are taken into account, (the formation of the lattice will be depressed below $H_{c2}^{mft}$, where amplitude of $\psi$ forms), it is generally believed that the transition will become 1st-order. [7]

Thermal fluctuations were first studied by BNT [7] in $\epsilon = 6 - d$ expansion about the upper critical dimension $d_{uc} = 6$. They found that fluctuations force one to consider an infinite set of marginally relevant quartic operators and derived a one-loop functional renormalization group recursion equation for the quartic coupling function. However, numerically analyzing these integro-differential equation, they were not able to find a fixed point function within the $\epsilon$-expansion. Consistent with their physical argument, BNT interpreted the runaway rg flows toward a negative coupling function as a fluctuation-driven 1st-order transition.

In clean samples the experiments observe a 1st-order melting transition of the Abrikosov vortex lattice [8], consistent with the conclusions of BNT. However, more recent experiments have found that this line of the 1st-order transition terminates at a critical end point, at a well-defined magnetic field, giving way to a 2nd-order transition [9].

In this Letter, I study the $H_{c2}$ NS transition by generalizing the GL theory to a large number $m$ of complex components of the superconducting order parameter $\psi_i$ and solve it in the $m \to \infty$ limit in arbitrary dimensionality. I find a 2nd-order transition in contrast to a general belief of a 1st-order transition, which, however only survives for $d > d_{lc} = 4$ and therefore predict that the superconducting phase coherence does not survive thermal fluctuations in physical dimensionality ($d = 2, 3$).

Near $H_{c2}$, considering strongly type-II superconductors allows me to ignore gauge field fluctuations. I take $\vec{A} = \frac{1}{2}H(-y, x; \mathbf{0}_\perp)$, as the symmetric gauge d-dimensional generalization of $H$ field perpendicular to the x-y plane. I solve the problem in the lowest Landau level (LLL) approximation valid near $H_{c2}$. In this case the quadratic part of the Hamiltonian Eq.1 can be diagonalized by $\psi_i(z, \bar{z}; \vec{r}_\perp) = \phi_i(z; \vec{r})e^{-|z|^2/4l^2}$, where $\phi_i(z; \vec{r})$ is an arbitrary $m$-component analytic function of $z = x + iy$ (not to be confused with the z coordinate), $\vec{r}_\perp$ denotes the $d - 2$-dimensional space perpendicular to the x-y plane defined by the $H$ field, and $l \equiv 1/\sqrt{qH}$ is the magnetic length. Substituting this form for $\psi_i$ into Eq.1 I obtain,

$$F[\phi_i] = \int d^{d_\perp}r_\perp d^2z_1 e^{-|z_1|^2/2l^2}\left[(|\vec{\nabla}_\perp \phi_i|^2 + t_o|\phi_i|^2) + \frac{1}{4}\int d^2z_2 e^{-|z_2|^2/2l^2} g_o(|z_1 - z_2|)|\phi_i(z_1, \vec{r}_\perp)|^2|\phi_j(z_2, \vec{r}_\perp)|^2\right], \quad (2)$$

where $t_o = r_o(T) + qH \propto (H - H_{c2}(T)) \propto (T - T_c(H))$ changes sign at the mean-field theory (MFT) transition and the interaction coupling constant $g_o$ has been generalized from $g_o \delta^{(2)}(z_1 - z_2)$ to an arbitrary short-range function $g_o(|z_1 - z_2|)$ evolution of which to a fixed point function (due to thermal renormalization) is the main result of this work. The fact that one must keep track of the whole function $g(|z|)$, rather than expanding can be seen as follows. In order to reach the NS transition fixed point, while $r_\perp$ is treated as usual distances rescaling under dilation by $b$ as, $r_\perp \to r_\perp b$, $z$ must remain unrescaled $z \to z$, in order to preserve under rescaling the form of the quadratic part of the free energy. If $z$ were rescaled with any power of $b$, the $e^{-|z|^2/2l^2}$ would reduce the free energy to zero, as $b \to \infty$, clearly an unsatisfactory situation. The nondimensionality of $z$ therefore leads to an infinite set of quartic coupling constants, labeled by $z$ and encoded into the $g(|z|)$ coupling function. Physically this behavior is due to the fact that while the correlation length $\xi_\perp$ diverges (I rescale $r_\perp$ to keep up with this divergence), the $q = 0$ wavevector correlations in the x-y plane grow but are eventually cutoff by the magnetic length $l$. All the long-wavelength modes labeled by $z$ are equally important in contributing to the divergences in $k_\perp \to 0$ susceptibility and other correlation functions, and therefore must be treated on equal footing. Rescaling of $z \to zb$ would amount to looking at the theory characterized by a vanishing magnetic length $l \to l/b$ or equivalently divergent $H$ field, a regime in which I do not expect a finite $T_c$ NS transition.

I now follow a standard large $m$ treatment [10], which surprisingly, can be easily generalized to the problem at hand. Via a Hubbard-Stratanovich transformation I introduce an auxiliary field $\chi(z, \vec{r}_\perp)$ that linearly couples to $|\phi_i(z_1, \vec{r}_\perp)|^2$ and has a bare propagator $g_o(|z_1 - z_2|)$ with the inverse defined by $\int d^2z g_o^{-1}(|z_1 - z|)g_o(|z - z_2|) = \delta^{(2)}(z_1 - z_2)$. Anticipating a breaking of symmetry along one of the $i$-directions, I take $\phi_i = (\sigma m, \pi_\alpha)$ and integrate over the $m - 1$ complex (Goldstone modes to-be) components of $\pi_\alpha$. For $g_o(|z_1 - z_2|)$ of order $1/m$, in the large $m$ limit, the fluctuations in the fields $\sigma$ and $\chi$ can be ignored (they lead to $1/m$ corrections) and I obtain the exact fluctuation corrected effective free energy

$$\frac{F[\sigma, \chi]}{m} = \int d^{d_\perp}r_\perp d^2z_1 \left[e^{-|z_1|^2/2l^2}\sigma^*\left(-\nabla_\perp^2 + t_o + \chi(z_1, \vec{r}_\perp)\right)\sigma - \frac{1}{m}\int d^2z_2 \ g_o^{-1}(|z_1 - z_2|)\chi(z_1, \vec{r}_\perp)\chi(z_2, \vec{r}_\perp)\right] + \left(1 - \frac{1}{m}\right)\text{Tr log}\left(-\nabla_\perp^2 + t_o + \chi(z, \vec{r}_\perp)\right), \quad (3)$$

To make analytical progress in evaluating the free energy, I will make the following choice for the low energy ansatz,

$$\sigma(z, \vec{r}_\perp) = \alpha_G \sigma_G(z) \equiv \alpha_G \prod_i (z - z_i),$$
$$\chi(z, \vec{r}_\perp) = \chi_o, \quad (4)$$

and now explain the physically motivation behind this choice. Obviously taking $\sigma(z, \vec{r}_\perp)$ to be uniform along $\vec{r}_\perp$ lowers the free energy. Since I want to describe the transition into an Abrikosov-type of phase, i.e. a phase of periodic arrangement of vortices characterized by $\sigma_G(z)$ (probably a triangular lattice given by a global minimization procedure), I take $\sigma(z)$ to be proportional to this a priori assumed low T phase. The complex amplitude coefficient $\alpha_G$ characterizes the level of order in this phase. Since the square of the superconducting order parameter $|\psi(z)|^2$, is essentially constant (aside from a lattice of zeros), $\int d^2z|\sigma(z)|^2e^{-|z|^2/2l^2} = Aa\alpha_G^2 = A\rho_s$ is extensive with the x-y area $A$ ($a$ is the reduction factor in the superfluid density $\rho_S^o$ due to zeros). Since $\chi(z, \vec{r}_\perp)$ couples to $|\sigma(z, \vec{r}_\perp)|^2$, with equivalent thermal averages, it is a good approximation to take it to be a constant, as I have done in Eq.4. Certainly the nonuniformity coming from the zeros in $\chi_o(z)$ will lead to numerical corrections but it is unlikely that they will result in any modification of the universal and qualitative properties of the NS transition.

Another important point is that I am obviously *assuming* the form of the low-T phase, drawing on the knowledge of a lattice solution. This is in the spirit of Landau's theory of freezing where the effective theory of the transition is for the slowly varying *coefficient* $\rho_G$ of the particular Fourier component of the full density at which the ordering is to occur. As in the theory of freezing, it is an exceptionally difficult task to analytically demonstrate a local instability to this finite wavevector $G$ ordering, which is brought about by the repulsive interactions. For the theory of freezing one does a separate difficult calculation within the disordered (liquid) phase to show that interactions lead to a peak in the structure function, approximately at the inverse of the interparticle separation, this peak being the precursor of the inevitable Bragg peak. The analogous task of showing (within the GL theory) a susceptibility to order at a finite in-plane wavevector, into for example the Abrikosov lattice, has not, to my knowledge, been performed analytically, even at the MFT level. Instead, one simply *assumes* a periodic solution and picks the global lowest energy one. Since the fluctuation-corrected free energy in Eq.3 is significantly different from the MFT one, it is possible, that upon minimization, it will lead to a lattice of a different form from the triangular one.

Having "apologized" for my choice of the ansatz, I now minimize $F$ in Eq.3 with respect to $\alpha_G$ and $\chi_o$, obtaining the saddle point equations

$$\alpha_G \xi^{-2} = 0, \quad (5)$$

$$\alpha_G - b(\xi^{-2} - t_o) + a^{-1} \int \frac{d^{d_\perp} k_\perp}{(2\pi)^{d_\perp}} \frac{1}{k_\perp^2 + \xi^{-2}} = 0, \quad (6)$$

where $b = 2/(ma) \int d^2 z g_o^{-1}(|z|)$, and I defined the correlation length $\xi$ by $\xi^{-2} = t_o + \chi_0$. For $T < T_c(H)$, $\alpha_G > 0$, and Eq.5 implies a divergent correlation length ($\xi^{-2} = 0$) in the Abrikosov phase. Eq.6 then leads to the order parameter $\alpha_G$, vanishing as a power-law with the reduced temperature, as the $T_c(H)$ is approached from below

$$\alpha_G \sim (t_c - t_o(H,T))^\beta \sim (T_c(H) - T)^\beta, \quad \beta = \frac{1}{2}, \quad (7)$$

where $t_c = -a^{-1} \int \frac{d^{d_\perp} k_\perp}{(2\pi)^{d_\perp}} \frac{1}{k_\perp^2}$ is the shift in $T_c(H)$ due to thermal fluctuations. I note that for $d = d_\perp + 2 < 4$ the thermal suppression of $T_c$ is divergent and therefore identify the lower-critical dimension as $d_{lc} = 4$. This is consistent with conclusions about $d_{lc}$ by Moore based on a completely different arguments applied in the ordered Abrikosov phase [11]. Eq.7 implies a vanishing of the superfluid density as $T_c(H)$ is approached from below as $\rho_s \sim |\alpha_G|^2 \sim (T_c(H) - T)$. For $T > T_c(H)$ the solution is characterized by a finite $\xi$. Eq.5 then implies $\alpha_G = 0$ and the state is normal. Eq.6 then describes how the correlation length $\xi$ diverges as $T_c(H)$ is approached from above

$$\xi \sim (T - T_c(H))^{-\nu}, \quad (8)$$

where for $d > d_{uc} = 6$ the MFT is accurate and $\nu = \frac{1}{2}$, while for $d_{lc} < d < d_{uc}$ thermal fluctuations are divergent and lead to $\nu = 1/(d-4)$. These are, as expected, the usual large $m$ exponents in the $d-2$ dimensional theory, with the dimensional reduction occuring due to the

$H$ field quenching of the kinetic energy in the x-y plane, perpendicular to it.

I now turn to the main focus of my work, that is, the effective free energy that describes the *fluctuations* about this large $m$ solution. In the large $m$ limit, the renormalized quartic interaction $g_R(|z_1 - z_2|)$ is the quantity that characterizes the fixed point of the NS transition. One way of computing it is to go back to the free energy in Eq.2 and to simply pertubatively resum all the loop diagrams to lowest order in $1/m$, which renormalize the bare quartic interaction $g_o(|z_1 - z_2|)$. Equivalently, given the formalism that I have set up above, I can simply compute the renormalization of the quadratic term in $\chi$ by expanding the free energy in Eq.3 to quadratic order in fluctuations of $\chi$ about the saddle point value $\chi_o$. Either approach gives,

$$g_R(q, k_\perp, t) = \frac{g_o(q)}{1 + m\Pi(q, k_\perp, t) g_o(q)}, \quad (9)$$

$$\Pi(q, k_\perp, t) = \frac{1}{4\pi l^2} I(k_\perp, t) e^{-|q|^2 l^2/2}, \quad (10)$$

where $q$ is a wavevector in the x-y plane, $\Pi(q, k_\perp, t)$ is the Fourier-transform of the polarization bubble with respect to complex $z$ (x-y) coordinate and

$$I(k_\perp, t) = \int \frac{d^{d_\perp} p_\perp}{(2\pi)^{d_\perp}} \frac{1}{(p_\perp^2 + t)(|\vec{p}_\perp - \vec{k}_\perp|^2 + t)}, \quad (11)$$

$$I(k_\perp, 0) = c(d) k_\perp^{-(6-d)}, \quad (12)$$

where in Eq.12 $I(k_\perp, t)$ is evaluated at $T_c$ ($t = 0$), and $c(d) = \Gamma(2 - d_\perp/2)\Gamma(d_\perp/2 - 1)^2/\Gamma(d_\perp - 2)/(4\pi)^{d_\perp/2}$, diverging as $1/(6-d)$ near $d_{uc}$, consistent with the $\epsilon$-expansion of BNT. The interaction function $g_R(q, k_\perp, t = 0)$, at $T_c$, is displayed in the inset of Fig.1 as a function of $q$, for a fixed small $k_\perp$, large $m$, and for a choice of short-range microscopic GL interaction $g_o(q) = e^{-a^2 q^2/2}$, with $a = l = 1$. It shows a peak at a finite wavevector $q = G \approx \text{Min}\{l^{-1}\sqrt{|\ln k_\perp|}, a^{-1}\}$, as one might expect for a continuous freezing transition, with a peak possibly being a precursor of the eventual Bragg singularities, not captured by the large $m$ theory. In real space $g_R(r_{xy}, k_\perp, t = 0)$ exhibits oscillations with a $1/r_{xy}^2$ power law fall off, reflecting correlations building up in the xy plane. Eqs.9-12 contain the information about the MFT fixed point (valid for $d > 6$), the new Heisenberg-type fixed point, characterizing the NS transition in the presence of thermal fluctuations, and about the crossover between them as a function of the reduced temperature $t$ and length scale $k_\perp$, labeled by the x-y wavevector $q$. Right at $T_c(H)$, for fixed $q$, $\Pi(q, k_\perp, 0) \sim k_\perp^{-\epsilon}$ diverges at long scales and MFT crosses over (with a crossover exponent $\phi = -\epsilon$) to the new fixed point summarized by Eqs.5-8, and with the effective interaction for fluctuations given by a completely *universal function*, (actually a distribution), independent of the bare interaction $g_o(|z_1 - z_2|)$

$$g_R(q, k_\perp, 0) = \frac{1}{m\Pi(q, k_\perp, 0)} = g k_\perp^{(6-d)} e^{|q|^2 l^2/2} , \quad (13)$$

$$g_R(|z_1 - z_2|, k_\perp, 0) = g k_\perp^{(6-d)} \delta^{(2)}(z_1 - z_2) \, e^{-\nabla_{xy}^2 l^2/2} , \quad (14)$$

where $g = 4\pi l^2/m/c(d)$. This form of $g_R$ is valid for arbitrary $d$, but for $6 - d = \epsilon << 1$ gives the fixed point function of BNT rg equations, in $m \to \infty$ limit [12].

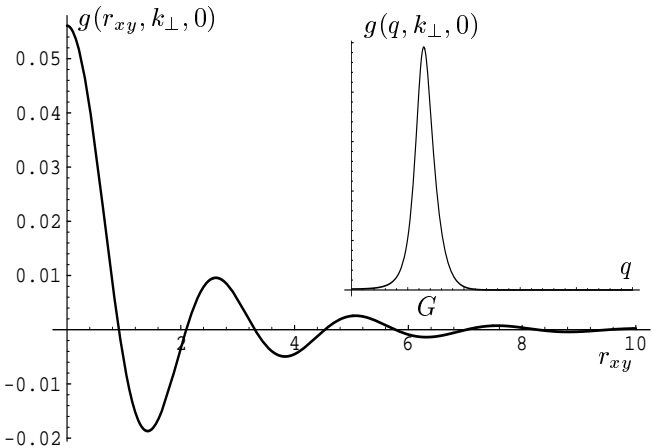

FIG. 1. $g_R(r_{xy}, k_\perp, t = 0)$ at $T_c$ for fixed $k_\perp$ exhibits density correlations in the xy plane manifested in oscillations at a scale $G^{-1}$; the inset shows $g_R(q, k_\perp, t = 0)$.

The fluctuations near $H_{c2}$ at the NS critical point are described by the effective free energy of same form as Eq.2, but with $g_o$ replaced by the *universal* function $g_R$ from Eq.14. Because $\chi$ couples to $|\psi|^2$, it is easy to show that its renormalized propagator $g_R$ is also the superfluid density susceptibility. This allows me to extract the specific heat $\alpha = -(6-d)/(d-4)$ and the correlation length $\nu = 1/(d-4)$ exponents, consistent with the value of $\nu$ obtained from the saddle point equations. Note that because of the unusual scaling of lengths at this critical point, discussed in the introduction, the standard relation $\alpha = 2 - d\nu$ is replaced by an easily derivable relation appropriate for this problem, $\alpha = 2 - (d-2)\nu$. The screening by fluctuations has removed the MFT specific heat divergence ($\alpha_{mft} = (6-d)/2 > 0$) and replaced it by a nondivergent nonanalyticity in $k_\perp$ with $\alpha < 0$. This suggests by a variant of the Harris criterion that the *infinitesimal* short-range point disorder is irrelevant for this finite field NS transition [12]. Similar arguments suggest that the transition is destabilized by the long-range disorder such as twin planes and the artificially introduced columnar defects [13].

Two of the important remaining problems are the demonstration of a diverging susceptibility to order into a lattice of vortices (i.e. divergent $\xi_{xy}$ at $q \sim G$), and the computation of the $1/m$-corrections. Preliminary calculations indicate that these problems might be related, and that the renormalized interaction in Eq.14 can lead to an instability to a vortex lattice, when the $1/m$ corrections to the effective free energy are computed [13].

In summary, I have presented a large $m$ theory of a 2nd-order $H_{c2}$ NS transition for arbitrary $d$ and found the fixed point effective free energy describing the transition. Since the physical dimensionalities $d = 2, 3$ are below the $d_{lc} = 4$, the transition does not survive thermal fluctuations in the limit of large $m$. It is, however, possible that for the physical value of $m = 1$, that the true lower-critical dimension will be reduced below $d = 3$, allowing the finite-T, *continuous* NS transition to persist.

Note added: After this work was submitted for publication I became aware of a interesting but unfortunately relatively unknown paper by I. Affleck and E. Brezin, Nucl. Phys. B **257**, 451 (1985) that also treats $H_{c2}$ transition in $m \to \infty$ limit. In contrast to my work, where I focus on the *fluctuation* part of the renormalized free energy, that work studies only the $m = \infty$ saddle point equations to the *constant* part of the free energy (my Eq.3). Affleck, et al. argue that there is no solution to the saddle point equations near $H_{c2}$, interpreting it as a 1st-order transition, in contrast to my conclusion. We have not been able to reconcile a possible absence of a saddle point solution and my unambiguous finding of the fixed point free energy (determined by Eq.14), usually indicative of a 2nd order transition.

I thank M. Moore for many enlightening discussions. I acknowledge the support by the NSF (DMR 94−16926), through the Science and Technology Center for Superconductivity. I am grateful to The Institute for Theoretical Physics at UCSB, where this work was done, for their hospitality, and their support under NSF Grant No. PHY89-04035.

[1] B. I. Halperin, T. C. Lubensky and S. K. Ma, Phys. Rev. Lett. **32**, 292 (1974).
[2] C. Dasgupta and B. I. Halperin, Phys. Rev. Lett. **47**, 1556 (1981)
[3] L. Radzihovsky, Euro. Phys. Lett. **29**, 227 (1995).
[4] A. A. Abrikosov, Zh. Eskp. Teor. Fiz **32**, 1442 (1957) [Sov. Phys. – JETP **5**, 1174 (1957)].
[5] P. Lee and S. Shenoy, Phys. Rev. Lett. **28**, 1025 (1972).
[6] A. Bray, Phys. Rev. B **9**, 4752 (1974); S. Hikami and A. Fujita, Phys. Rev. B **41**, 6379 (1990).
[7] E. Brezin, D. R. Nelson and A. Thiaville, Phys. Rev. B **31**, 7124 (1985).
[8] H. Safar *et al.*, Phys. Rev. Lett. **69**, 824 (1992).
[9] H. Safar *et al.*, Phys. Rev. Lett. **70**, 3800 (1993).
[10] J. Zinn-Justin *Quantum Field Theory and Critical Phenomena* (Oxford University Press, NY, 1989).
[11] M. A. Moore, Phys. Rev. B **45**, 7336 (1992).
[12] M. A. Moore, T. Newman and L. Radzihovsky, unpublished.
[13] L. Radzihovsky, unpublished.